\documentclass{elsart}

\newif\ifpdf
\ifx\pdfoutput\undefined
    \pdffalse
\else
    \pdfoutput=1
    \pdftrue
\fi

\usepackage{subeqnarray}

\ifpdf
    \usepackage[pdftex]{graphicx}
    \usepackage[pdftex]{hyperref}
    \hypersetup{
        pdftitle = {Population trapping in the one-photon mazer},
        pdfauthor = {T. Bastin, E. Solano},
        pdfsubject = {Publication},
        pdfkeywords = {Quantum optics}
    }
    \DeclareGraphicsExtensions{.pdf}
\else
    \usepackage[dvips]{graphicx}
    \usepackage{hyperref}
    \DeclareGraphicsExtensions{.eps}
\fi

\begin{document}

\begin{frontmatter}

\title{Population trapping in the one-photon mazer}

\author{T. Bastin}
\address{Institut de Physique Nucl\'eaire, Atomique et de Spectroscopie, Universit\'e
  de Li\`ege au Sart Tilman, B\^at. B15, B - 4000 Li\`ege, Belgique}

\author{E. Solano}
\address{Max-Planck-Institut f\"ur Quantenoptik, Hans-Kopfermann-Strasse 1,
D-85748 Garching, Germany}
\address{Secci\'{o}n F\'{\i}sica, Departamento de
Ciencias, Pontificia Universidad Cat\'{o}lica del Per\'{u},
Apartado 1761, Lima, Peru}

\date{Received: \today / Revised version: date}

\begin{abstract}
We study the population trapping phenomenon for the one-photon
mazer. With this intent, the mazer theory is written using the
dressed-state coordinate formalism, simplifying the expressions
for the atomic populations, the cavity photon statistics, and the
reflection and transmission probabilities. Under the population
trapping condition, evidence for new properties of the atomic
scattering is given. Experimental issues and possible applications
are discussed.
\end{abstract}

\end{frontmatter}

{PACS number~: 42.50.-p, 32.80.-t, 42.50.Ct, 42.50.Dv}

\section{Introduction}
Laser cooling of atoms is a rapidly developing field in quantum
optics. Cold and ultracold atoms introduce new regimes in atomic
physics often not considered in the past. In the last decade,
Englert \etal~\cite{Eng91} have demonstrated new interesting
properties in the interaction of cold atoms with a micromaser
field. More recently, Scully \etal~\cite{Scu96} have shown that a
new kind of induced emission occurs when a micromaser is pumped by
ultracold atoms, requiring a quantum-mechanical treatment of the
center-of-mass motion. They called this particular process mazer
action to insist on the quantized $z$-motion feature of the
induced emission.

The detailed quantum theory of the mazer has been presented in a
series of three papers by Scully and co-workers
\cite{Mey97,Lof97,Sch97}. They showed that the induced emission
probability is strongly dependent on the cavity mode profile.
Analytical calculations were presented for the mesa and the
sech$^2$ mode profiles. For sinusoidal modes, WKB solutions were
detailed.

Retamal \etal~\cite{Ret98} showed that we must go beyond the WKB
solutions for the sinusoidal mode case when we consider strictly
the ultracold regime. Remarkably, they showed that the resonances
in the emission probability are not completely smeared out for
actual interaction and cavity parameters. In a recent work
\cite{Bas00}, we proposed a numerical method for calculating
efficiently the induced emission probability for arbitrary cavity
field modes. In particular, the gaussian potential was considered,
thinking in open cavities in the microwave or optical field
regime. Differences with respect to the sech$^2$ mode case were
found. Calculations for sinusoidal potentials were also performed
and divergences with WKB results were reported, confirming results
given in \cite{Ret98}.

Zhang \etal~\cite{Zha99} extended the concept of the mazer to the
two-photon process by proposing the idea of the two-photon mazer.
Their work was focused on the study of its induced emission
probability in the special case of the mesa mode function. Under
the condition of an initial coherent field state, they showed that
this probability exhibits with respect to the interaction length
the collapse and revival phenomenon, which have different features
in different regimes. They are similar to those in the two-photon
Jaynes-Cummings model only in the thermal-atom regime. Recently,
Arun \etal~\cite{Aru00} studied the mazer action in a bimodal
cavity with particular incidence in the mode-mode correlations.

The collapses and revivals of the atomic excitation in the
framework of the Jaynes-Cummings model was predicted in the early
1980s by Eberly and co-workers \cite{Ebe80,Nar81,Yoo81}. It is now
well established (see {\it e.g.} Ref.~\cite{Mey92} and references
therein) that this phenomenon is a direct consequence of the
interference of quantum Rabi floppings at various frequencies and
of the granularity of the field. Experimental evidence for
collapses and revivals has been reported by Brune
\etal~\cite{Bru96b} and Rempe \etal~\cite{Rem87} by use of a
micromaser device. Fleischhauer and Schleich \cite{Fle93} showed
later that the shape of each revival is a direct reflection of the
shape of the initial photon-number distribution $P_n$, assuming
that the atom is prepared completely in the upper state or in the
lower state and that the distribution $P_n$ is sufficiently
smooth. It was also noticed that, under some special conditions of
the initial atom-field state, the revivals can be largely and even
completely suppressed \cite{Zah89,Cir90,Gea91}. This phenomenon
was denominated ``population trapping'' to refer, as noted by Yoo
and Eberly \cite{Yoo85}, to a persistent probability of finding
the atom in a given level in spite of the existence of both the
radiation field and allowed transitions to other levels. The
initial atom-field states giving rise to this phenomenon were
called ``trapping states'' in \cite{Cir91}. Let us mention that
this denomination is actually used in various physical contexts
whenever a degree of freedom is found unaltered in spite of the
existence of an interaction able to change its value. For instance
trapping states in the context of the micromaser theory have been
predicted and very recently measured by Filipowicz
\etal~\cite{Fil86} and Weidinger \etal~\cite{Wei99} respectively.
Nevertheless, these trapping states do not relate with those
responsible for the suppression of the revivals mentioned in
\cite{Zah89,Cir90,Gea91}.

An elegant explanation of the population trapping phenomenon has
been proposed by Jonathan \etal~\cite{Jon99}, who noticed that the
key to understand the collapse and revival patterns under very
general conditions is to consider the joint initial properties of
the atom-field system, even if this one is completely disentangled
before the interaction. By defining an appropriate coordinate
system, the dressed-state coordinates, they were able to yield
simple analytical expressions for the atomic populations which
exhibit the conditions needed for population trapping.

As the $z-$motion quantization introduces a new kind of induced
emission dependent of the cavity mode profile, and thus strongly
modifies the usual Jaynes-Cummings atom-field evolution, we study
in this paper how the population trapping phenomenon is affected
by this motion quantization and we show that the trapping states
provide interesting new features to the mazer.

In Sec.~\ref{TheModelSection}, we write the quantum theory of the
one-photon mazer by use of the dressed-state coordinate formalism
as it was very efficient in the description of the population
trapping phenomenon in the Jaynes-Cummings model~\cite{Jon99}.
General expressions are derived for the atomic populations and the
cavity photon distribution after the interaction of the atom with
the cavity. The theory is written for any initial pure state of
the atom-field system (entangled or not). We consider zero
temperature and no dissipation in the high-$Q$ cavity.
Sec.~\ref{PopulationTrappingSection} is devoted to the population
trapping phenomenon for the one-photon mazer and new properties of
the trapping states in the scattering process are presented.
Experimental issues are briefly discussed in
Sec.~\ref{ExperimentalIssues}. A brief summary of our results is
given in Sec.~\ref{SummarySection}.

\section{The model}
\label{TheModelSection}
\subsection{The Hamiltonian}
We consider a two-level atom moving along the $z$-direction in the
way to a cavity of length $L$. The atom is coupled resonantly with
an one-photon transition to a single mode of the quantized field
present in the cavity. The atom-field interaction is modulated by
the cavity field mode function. The atomic center-of-mass motion
is described quantum mechanically and the rotating-wave
approximation is made. In the interaction picture, the Hamiltonian
describing the system is
\begin{equation}
\label{Hamiltonian} H = \frac{p^2}{2M}+ \hbar g \, u(z)
(a^{\dagger} \sigma + a \sigma^{\dagger}),
\end{equation}
where $p$ is the atomic center-of-mass momentum along the
$z$-axis, $M$ is the atomic mass, $\sigma = |b \rangle \langle a|$
($|a\rangle$ and $|b\rangle$ are respectively the upper and lower
levels of the atomic transition), $a$ and $a^{\dagger}$ are
respectively the annihilation and creation operators of the cavity
radiation field, $g$ is the atom-field coupling strength and
$u(z)$ is the cavity field mode.

\subsection{The wavefunctions}
In the $z$-representation and in the dressed-state basis
\begin{equation}
\label{DefDressedStateBasis} \left \{
\begin{array}{l}
|b,0 \rangle, \\ |\pm,n \rangle = \frac{1}{\sqrt{2}}\left(|a,n
\rangle \pm |b,n+1 \rangle\right),
\end{array}
\right.
\end{equation}
$|n\rangle$ being the photon-number states, the problem reduces to
the scattering of the atom upon the potentials $V_n^{\pm}(z) = \pm
\hbar g \sqrt{n+1} \, u(z)$ (see Ref.~\cite{Mey97}). Indeed, the
set of wavefunction components
\begin{equation}
\label{DefComponents} \psi_n^{\pm}(z,t)= \langle z,\pm,n | \psi(t)
\rangle ,
\end{equation}
where $|\psi(t)\rangle$ is the atom-field state satisfy the
Schr\"{o}dinger equation
\begin{equation}
\label{TimeDependentSchrodingerEquation} i \hbar
\frac{\partial}{\partial t} \psi_n^{\pm}(z,t) = \left( -
\frac{\hbar^2}{2M}\frac{\partial^2}{\partial z^2} +
V_n^{\pm}(z)\right) \psi_n^{\pm}(z,t) .
\end{equation}

The general solution of (\ref{TimeDependentSchrodingerEquation})
is
\begin{equation}
\label{GeneralSolution} \psi_n^{\pm}(z,t) = \int \! dk \,
\phi_n^{\pm}(k) e^{-i \frac{\hbar k^2}{2M} t} \varphi_n^{\pm}(k,z)
,
\end{equation}
where $\varphi_n^{\pm}(k,z)$ is solution of the time-independent
Schr\"{o}dinger equation
\begin{equation}
\label{TimeIndependentSchrodingerEquation} \left(
\frac{\partial^2}{\partial z^2} + k^2 \mp \kappa_n^2 u(z) \right)
\varphi_n^{\pm}(k,z) = 0 ,
\end{equation}
with
\begin{equation}
\label{kappanDef} \kappa_n = \kappa \sqrt[4]{n+1}
\end{equation}
and
\begin{equation}
\label{kappaDef} \kappa = \sqrt{2M g/\hbar} \: .
\end{equation}

The wavefunction component
\begin{equation}
\label{psimnzt} \psi_{-1}(z,t) = \langle z, b, 0 | \psi(t) \rangle
\end{equation}
satisfy a Schr\"{o}dinger equation characterized with a null
potential and is therefore not affected by the interaction of the
atom with the cavity. The atom in the lower state cannot obviously
interact with the cavity field that does not contain any photon.
The component (\ref{psimnzt}) describes a free particle problem.

We assume that, initially, the atomic center-of-mass motion is not
correlated to the other degrees of freedom. We describe it by the
wave packet
\begin{equation}
\label{InitialStateZ} \chi(z) \equiv \langle z | \chi \rangle =
\int \! dk A(k) e^{ikz}\theta(-z) ,
\end{equation}
where $\theta(z)$ is the Heaviside step function (indicating that
the atoms are incident from the left of the cavity). No
restrictions are made for the initial conditions of the atomic
internal state and the cavity field state, except that pure states
are only considered. By use of an expansion over the dressed-state
basis (\ref{DefDressedStateBasis}), we may write
\begin{equation}
\label{psi0preamble} | \psi(0) \rangle \!=\!\! | \chi \rangle
\otimes \! \left( w_{-1}e^{i\chi_{-1}}|b,0\rangle \!+\!\!\!
\sum_{n = 0}^{\infty} w_{n}e^{i\chi_{n}} | \beta_n \rangle
\right),
\end{equation}
with
\begin{equation}
| \beta_n \rangle = \cos\left(\frac{\theta_n}{2}\right)|+,n\rangle
+ e^{-i\phi_n}\sin\left(\frac{\theta_n}{2}\right)|-,n\rangle .
\end{equation}

The parameters $w_n \in [0,1]$, $\theta_n \in [0,\pi]$ and
$\chi_n$, $\phi_n \in [0,2\pi]$ are called dressed-state
coordinates \cite{Jon99}. The normalisation condition is
\begin{equation}
\sum_{n=-1}^{\infty}w_n^2 = 1
\end{equation}
and the phase factor $\chi_{-1}$ may be set to $0$ without loss of
generality.

We consider therefore
\begin{equation}
\left \{
\begin{array}{l}
\psi_{-1}(z,0) = c_{-1} \chi(z) , \\ \psi_n^{\pm}(z,0) =
c^{\pm}_{n} \chi(z) ,
\end{array}
\right.
\end{equation}
with
\begin{equation}
\left \{
\begin{array}{l}
c_{-1} = w_{-1} ,\\ c^{+}_n = w_n e^{i \chi_n}
\cos\left(\theta_n/2\right) ,\\ c^{-}_n = w_n e^{i (\chi_n -
\phi_n)} \sin\left(\theta_n/2\right) .
\end{array}
\right.
\end{equation}

Inserting Eqs.~(\ref{DefDressedStateBasis}) and
(\ref{InitialStateZ}) into Eq.~(\ref{psi0preamble}), we get
\begin{eqnarray}
\label{psi0} |\psi(0) \rangle & = & \int \!\! dz \! \int \!\! dk
\, A(k) \times \nonumber \\ & & \Bigg( \sum_{n=0}^{\infty} \Big[
S_{a,n} e^{ikz}\theta(-z) |z,a,n \rangle \nonumber \\ & & + S_{b,n
+ 1} e^{ikz}\theta(-z) |z,b,n+1 \rangle \Big] \nonumber \\ & & +
w_{-1} e^{ikz}\theta(-z) |z,b,0 \rangle \Bigg) ,
\end{eqnarray}
with
\begin{equation}
\label{SanSbn}
\left(
\begin{array}{c}
S_{a,n} \\ S_{b,n+1}
\end{array}
\right)= \tilde{A}_n \left(\begin{array}{c} 1 \\
1\end{array}\right)
\end{equation}
and
\begin{equation}
\tilde{A}_n = \frac{w_n e^{i\chi_n}}{\sqrt{2}}
\left(\begin{array}{cc} \cos\left(\theta_n/2\right) & e^{-i\phi_n}
\sin\left(\theta_n/2\right) ,\\ \cos\left(\theta_n/2\right) &
-e^{-i\phi_n} \sin\left(\theta_n/2\right) ,
\end{array}
\right)
\end{equation}

After the atom has left the interaction region, the wavefunctions
$\varphi_n^{\pm}(k,z)$ can be written as
\begin{equation}
\label{varphi} \varphi_n^{\pm}(k,z) = \left \{ \begin{array}{ll}
r_n^{\pm}(k)e^{-ikz} & (z < 0) \\ t_n^{\pm}(k)e^{ik(z-L)} & (z >
L)
\end{array}\right. ,
\end{equation}
where $r_n^{\pm}(k)$ and $t_n^{\pm}(k)$ are respectively the
reflection and transmission coefficient associated with the
scattering of the particle of momentum $\hbar k$ upon the
potential $V_n^{\pm}(z)$
(Eq.~\ref{TimeIndependentSchrodingerEquation}). The initial state
components $\psi_n^{\pm}(z,0)$ have evolved into
\begin{eqnarray}
\psi_n^{\pm}(z,t) & = & c^{\pm}_n \int \! dk \, A(k) e^{-i
\frac{\hbar k^2}{2M} t} \big[r_n^{\pm}(k)e^{-ikz}\theta(-z)
\nonumber \\ & & + t_n^{\pm}(k)e^{ik(z - L)}\theta(z - L) \big]
\end{eqnarray}
whereas the free particle wavefunction component $\psi_{-1}(z,0)$
becomes
\begin{equation}
\psi_{-1}(z,t) = c_{-1} \int \! dk \, A(k) e^{-i \frac{\hbar
k^2}{2M} t} e^{ik(z - L)} \theta(z - L) .
\end{equation}

We thus obtain
\begin{eqnarray}
\label{psit} |\psi(t) \rangle & = & \int \!\! dz \! \int \!\! dk
\, A(k) e^{-i \frac{\hbar k^2}{2M} t} \times \nonumber \\ & &
\Bigg( \sum_{n=0}^{\infty} \Big[ R_{a,n}(k) e^{-ikz} \theta(-z)
|z,a,n \rangle \nonumber \\ & &     + T_{a,n}(k) e^{ik(z-L)}
\theta(z-L)   |z,a,n \rangle \\ & &     + R_{b,n+1}(k) e^{-ikz}
\theta(-z) |z,b,n+1 \rangle \nonumber \\ & &     + T_{b,n+1}(k)
e^{ik(z-L)} \theta(z-L) |z,b,n+1 \rangle \Big] \nonumber \\ & & +
w_{-1} e^{ik(z-L)} \theta(z-L) |z,b,0 \rangle \Bigg) , \nonumber
\end{eqnarray}
in which
\begin{subeqnarray}
\label{RanRbnTanTbn}
\left(\begin{array}{c} R_{a,n}(k) \\
R_{b,n+1}(k)
\end{array}
\right) & = & \tilde{A}_n \left(
\begin{array}{c} r_n^+(k) \\ r_n^-(k)
\end{array} \right) ,
\slabel{RanRbn} \\
 \left(
\begin{array}{c}
T_{a,n}(k) \\ T_{b,n+1}(k)
\end{array}
\right) & = & \tilde{A}_n \left(
\begin{array}{c}
t_n^+(k) \\ t_n^-(k)
\end{array}
\right) . \slabel{TanTbn}
\end{subeqnarray}

If initially the electromagnetic field is in the state $|n
\rangle$ and the atom is in the excited state~$|a \rangle$, the
only non-zero dressed-state coordinates are $w_n = 1$ and
$\theta_{n} = \pi / 2 $. We get therefore
\begin{equation}
\tilde{A}_n = \frac{1}{2} \left(
\begin{array}{cc}
1 & 1 \\ 1 & -1
\end{array}
\right)
\end{equation}
and Eqs.~(\ref{RanRbnTanTbn}) lead to the same results given by
Meyer \etal~\cite{Mey97} who considered in detail this case for
the one-photon mazer.

\subsection{Atomic populations}
The reduced density matrix $\sigma(t)$ for the atomic internal
degree of freedom is given by the trace over the radiation and the
atomic external variables of the atom-field density matrix, that
is its elements $i,j = a,b$ are
\begin{equation}
\label{sigmat} \sigma_{ij}(t) = \sum_{n} \int \!\! dz \langle
z,i,n | \psi(t) \rangle \langle \psi(t) | z,j,n \rangle .
\end{equation}
The atomic populations $\sigma_{ii}$ follows immediately from
Eq.~({\ref{sigmat})~:
\begin{equation}
\label{sigmaiit} \sigma_{ii}(t) = \sum_{n} \int \!\! dz |\langle
z,i,n | \psi(t) \rangle |^2 .
\end{equation}

Inserting Eqs.~(\ref{psi0}) and (\ref{psit}) into
Eq.~(\ref{sigmaiit}) and using Eqs.~(\ref{SanSbn}) and
(\ref{RanRbnTanTbn}), we get for an incident atom of momentum
$\hbar k$~:
\begin{eqnarray}
\label{sigmaAAInitial} \sigma_{aa}(0) & = & \frac{1}{2}\left[ 1 -
w_{-1}^2 + \sum_{n=0}^{\infty} w_n^2 \sin(\theta_n) \cos(\phi_n)
\right] ,\\ \label{sigmaAAt} \sigma_{aa}(t) & = &
\frac{1}{2}\left[ 1 - w_{-1}^2 + \sum_{n=0}^{\infty} w_n^2
\sin(\theta_n) \textrm{Re}(e^{i\phi_n}
K_n) \right] ,\nonumber \\
\end{eqnarray}
where
\begin{equation}
\label{defKn} K_n = r^+_n r^{-*}_n + t^+_n t^{-*}_n .
\end{equation}

The change of the atomic population $\sigma_{aa}$ induced by the
interaction of the incident atom with the cavity radiation field
is then given by
\begin{equation}
\delta \sigma_{aa} = \sigma_{aa}(t) - \sigma_{aa}(0) ,
\end{equation}
with the time $t$ chosen long after the interaction.

Thus we have
\begin{equation}
\label{deltaSigmaAA} \delta \sigma_{aa} = \sum_{n=0}^{\infty}
\Delta_n ,
\end{equation}
with
\begin{equation}
\label{deltaN} \Delta_n = \frac{w_n^2}{2} \sin(\theta_n) \left[
\textrm{Re} \left( e^{i\phi_n} K_n \right) - \cos(\phi_n) \right].
\end{equation}

As expected, the component $w_{-1}$ of the initial state
$|\psi(0)\rangle$ over the state $|b,0 \rangle$ does not play any
role in the dynamics of the system.

We have to emphasize that in Eq.~(\ref{deltaSigmaAA}) $\Delta_n$
\emph{cannot} be interpreted strictly as the change in the
$\sigma_{aa}$ population induced by the interaction of the
two-level atom with the cavity radiation field containing $n$
photons. This is only true when the incident atom is prepared in
the excited state. Indeed, if initially the internal atomic state
is $c_a |a\rangle + c_b |b\rangle$ and the field state is
$|n\rangle$ $(n \geq 1)$, then the only non-zero dressed-state
coordinates are $w_n = |c_a|$, $\chi_n = \arg(c_a)$, $\theta_n =
\pi/2$, $w_{n-1} = |c_b|$, $\chi_{n-1} = \arg(c_b)$, $\theta_{n-1}
= \pi/2$ and $\phi_{n-1} = \pi$. We thus have in that case
\begin{eqnarray}
\delta \sigma_{aa}  & = & \Delta_n + \Delta_{n-1} ,\nonumber \\ &
= & \Delta_n \quad \textrm{iff} \quad c_b=0.
\end{eqnarray}

\subsection{Photon statistics}
The reduced density matrix $\rho(t)$ for the cavity radiation
field is given by the trace over the internal and external atomic
degrees of freedom of the atom-field density matrix, that is its
elements $n,n'$ are
\begin{equation}
\label{rhot} \rho_{nn'}(t) = \sum_{i = a,b} \int \!\! dz \langle
z,i,n | \psi(t) \rangle \langle \psi(t) | z,i,n' \rangle .
\end{equation}

The photon distribution $P_n=\rho_{nn}$ follows immediately from
Eq.~({\ref{rhot})~:
\begin{equation}
\label{Pn} P_n(t) = \sum_{i = a,b} \int \!\! dz |\langle z,i,n |
\psi(t) \rangle |^2 .
\end{equation}

The change $\delta P_n$ in the cavity photon distribution induced
by the interaction of the cavity electromagnetic field with the
incident atom is then given by
\begin{equation}
\delta P_n = P_n(t) - P_n(0) .
\end{equation}

Inserting Eqs.~(\ref{psi0}) and (\ref{psit}) into Eq.~(\ref{Pn})
and using Eqs.~(\ref{SanSbn}) and (\ref{RanRbnTanTbn}), we get for
an incident atom of momentum $\hbar k$~:
\begin{equation}
\label{deltaPn} \delta P_n = \left \{
\begin{array}{ll}
\Delta_n -\Delta_{n-1} & (n \geq 1) , \\ \Delta_n & (n = 0) .
\end{array} \right.
\end{equation}

We see that if the initial state is $|a,n\rangle$ we have
\begin{equation}
\delta \sigma_{aa} + \delta P_{n + 1} = 0 ,
\end{equation}
which gives an intuitive population conservation condition.

\subsection{Reflection and transmission probabilities}
The reflection and transmission probabilities of the incident atom
upon the cavity are respectively given by
\begin{subeqnarray}
\label{defRandT} R & = & \sum_{i = a,b} \sum_{n}
\int_{-\infty}^{0} \!\! dz |\langle z,i,n | \psi(t) \rangle |^2 ,
\slabel{defR} \\ T & = & \sum_{i = a,b} \sum_{n} \int_{L}^{\infty}
\!\! dz |\langle z,i,n | \psi(t) \rangle |^2 . \slabel{defT}
\end{subeqnarray}

Inserting Eq.~(\ref{psit}) into Eqs.~(\ref{defRandT}), we get for
an incident atom of momentum $\hbar k$~:
\begin{subeqnarray}
\label{RandTExpr} R & = & \sum_{n=0}^{\infty} w_n^2 \Big(
\cos^2(\theta_n /2) |r_n^+|^2 + \sin^2(\theta_n /2) |r_n^-|^2
\Big) ,\\T & = & \sum_{n=0}^{\infty} w_n^2 \Big( \cos^2(\theta_n
/2) |t_n^+|^2 + \sin^2(\theta_n /2) |t_n^-|^2 \Big) \nonumber \\ &
& + w_{-1}^2 .
\end{subeqnarray}

One verifies immediately that the results of Meyer
\etal~\cite{Mey97} about the reflection and transmission
probabilities are well recovered by Eqs.~(\ref{RandTExpr}) when
their initial conditions are considered. Indeed, when the
atom-field system is initially in the state $|a,n\rangle$,
Eqs.~(\ref{RandTExpr}) become
\begin{subeqnarray}
\label{RandTExprSystInAn} R & = & \frac{1}{2} \big(|r^+_n|^2 +
|r^-_n|^2 \big) ,\\ T & = & \frac{1}{2} \big(|t^+_n|^2 + |t^-_n|^2
\big) .
\end{subeqnarray}

We get the same results if the atom-field system is initially in
the state $|b,n\rangle$ with $n \geq 1$, except that $n$ must be
replaced by $n - 1$ in Eqs.~(\ref{RandTExprSystInAn}). In the case
$n = 0$, we have obviously $T = 1$.

\subsection{Final remarks}
All the results here above (about the atomic populations, the
photon statistics, and the reflection and transmission
probabilities) may be very easily generalized for any momentum
wavefunction $A(k)$ of the initial wave packet. The various
expressions must simply be weighted by $|A(k)|^2$ and integrated
over $k$. For instance, Eq.~(\ref{deltaSigmaAA}) becomes
\begin{equation}
\label{deltaSigmaAAWithWavePacket} \delta \sigma_{aa} = \int \!\!
dk \, |A(k)|^2 \sum_{n=0}^{\infty} \Delta_n ,
\end{equation}
where $\Delta_n$ depends on $k$ through the reflection and
transmission coefficients, $r_n^{\pm}(k)$ and $t_n^{\pm}(k)$
respectively, in $K_n$ (see Eq.~(\ref{deltaN})). The expressions
obtained for all these various physical quantities are very simple
in the framework of the dressed-state formalism, even though they
are very general. They take a form much more complicated when the
usual coordinates of the atom-field system are used (the complex
coefficients $c_a$, $c_b$ and $c(n)$ of the atom-field states
written as $(c_a |a\rangle + c_b |b\rangle) \otimes \sum_n
c(n)|n\rangle$). Also entangled initial states may be considered
by this formalism. The great advantage of the dressed-state
coordinates was already pointed out by Jonathan \etal~\cite{Jon99}
who used them to express various physical quantities in the
Jaynes-Cummings model.

\section{Population Trapping}
\label{PopulationTrappingSection}

When the atom-field initial state is such that $\sin(\theta_n)=0$,
we get from Eq.~(\ref{deltaN}) $\Delta_n = 0$, whatever the value
of $K_n$. In this case, we have
\begin{equation}
\delta \sigma_{aa} = \delta \sigma_{bb} = \delta P_n = 0 ,
\end{equation}
indicating that the interaction of the atom with the cavity
radiation field has no effect on the atomic populations
$\sigma_{ii}$ ($i = a,b$) and on the cavity photon distribution
$P_n$, whatever the cavity field mode function, whatever the
cavity interaction length $\kappa L$ and whatever the atomic
initial velocity. We conclude that the mazer give rise to the
perfect population trapping phenomenon, when considering zero
temperature and no dissipation in the high-$Q$ cavity. This
property holds for the ultracold, intermediate and thermal-atom
regimes, as it is completely independent on the external atomic
degree of freedom. For the same reason, it holds for any momentum
wavefunction $A(k)$ of the initial wave packet.

The class of states verifying $\sin(\theta_n)=0$, named
\emph{perfect trapping states}, are given by (see
Ref.~\cite{Jon99})
\begin{equation}
| \gamma^{\pm} \rangle = \frac{ \gamma |a\rangle \pm |b\rangle
}{\sqrt{1 + |\gamma|^{2}}} \otimes \sqrt{1 - |\gamma|^2}
\sum_{n=0}^{\infty} \gamma^n |n\rangle ,
\end{equation}
where $\gamma$ is a complex number with $|\gamma| < 1$.

Indeed, rewriting these states in terms of the dressed-state
basis, we find
\begin{equation}
\label{gamma} | \gamma ^{\pm} \rangle = \sqrt{\frac{1 -
|\gamma|^2}{1 + |\gamma|^{2}}} \left( \sum_{n=0}^{\infty} \sqrt{2}
\gamma^{n+1} |\pm,n\rangle \pm |b,0\rangle \right) .
\end{equation}

For each $n$ there is only a single dressed-state present in the
sum of expression (\ref{gamma}). Depending on whether it is
$|+,n\rangle$ or $|-,n\rangle$, we have respectively
$\sin(\theta_n/2) = 0$ or $\cos(\theta_n/2) = 0$, and so
$\sin(\theta_n) = 0$ in any case. This give rise to another very
interesting feature of the perfect trapping states. The reflection
and transmission probabilities (\ref{RandTExpr}) become

\begin{subeqnarray}
\label{RandTtrap} R & = & \sum_{n=0}^{\infty} w_n^2 |r_n^{\pm}|^2
,\\ T & = & \sum_{n=0}^{\infty} w_n^2 |t_n^{\pm}|^2 + w_{-1}^2 ,
\end{subeqnarray}
with
\begin{subeqnarray}
\label{wnwm1} w_{n}  & = & \sqrt{\frac{1 - |\gamma|^2}{1 +
|\gamma|^{2}}} \sqrt{2} |\gamma|^{n+1} , \\ w_{-1} & = &
\sqrt{\frac{1 - |\gamma|^2}{1 + |\gamma|^{2}}} .
\end{subeqnarray}

Remarkably, the particle moving along the $z-$axis is only
sensitive to either a superposition of the potentials $V_n^+(z)$
(for the $|\gamma^+\rangle$ states) or a superposition of
$V_n^-(z)$ (for the $|\gamma^-\rangle$ states), but never to both.
So, it is possible to imagine an experimental set-up where the
particles would encounter only an effective potential well or only
an effective potential barrier, inhibiting the simultaneous action
of them.

Another important characteristic is that the perfect trapping
states do not make the cavity transparent to the incident atoms,
as in the conventional micromaser, because the reflection
coefficient $R$ is not nullified. In this way, we may say that
while there is no change in the atomic and field populations, the
quantization of the $z-$motion leads to an observable mechanical
effect of the atom-field interaction under the perfect trapping
condition.

Inserting (\ref{wnwm1}) into (\ref{RandTtrap}), we get
\begin{subeqnarray}
R & = & 2 \frac{1 - |\gamma|^2}{1 + |\gamma|^2} |\gamma|^2 \sum_{n=0}^{\infty} |\gamma|^{2n} |r_n^{\pm}|^2 ,\slabel{R1}\\
T & = & 2 \frac{1 - |\gamma|^2}{1 + |\gamma|^2} |\gamma|^2 \left(
\sum_{n=0}^{\infty} |\gamma|^{2n} |t_n^{\pm}|^2 + \frac{1}{2}
|\gamma|^{-2} \right). \slabel{T1}
\end{subeqnarray}

In the ultracold regime ($k \ll \kappa_n$), the transmission
probability through the potential barrier $V_n^+(z)$ is negligible
and we may consider $|t_n^+| = 0$ and $|r_n^+| = 1$, whatever the
cavity field mode. We thus have for the $|\gamma^+\rangle$ state
\begin{subeqnarray}
\label{RandTsimplified}
R & = & \frac{2 |\gamma|^2}{1 + |\gamma|^2} , \slabel{Rsimplified}\\
T & = & \frac{1 - |\gamma|^2}{1 + |\gamma|^2}.
\end{subeqnarray}

When $|\gamma| \rightarrow 0$, we have $R \rightarrow 0$ and the
cavity acts as a transmitter system. Inversely, when $|\gamma|
\rightarrow 1$, we get $R \rightarrow 1$ and the cavity acts as a
reflector system. By varying $|\gamma|$ between 0 and 1, we tune
the mazer from a perfect transmitter to a perfect reflector. In
both case, the atom-cavity interaction does not perturb in any way
the internal state of the system. This interesting property may be
easily understood like this. In the state $|\gamma^+\rangle$, the
ratio between the wavefunction components over the states
$|b,0\rangle$ and $|+,n\rangle$ is given by (see
eq.~(\ref{gamma}))
\begin{equation}
    \label{ratio}
    \frac{\omega_{-1}}{\omega_n} = \frac{1}{\sqrt{2}|\gamma|^{n +
    1}} .
\end{equation}

As $|\gamma|$ decreases, the wavefunction component over the state
$|b,0\rangle$ increases and this state does not give rise to any
interaction between the atom motion and the cavity. No reflection
may occur in this case. Inversely, as $|\gamma|$ increases, the
wavefunction components over the states $|+,n\rangle$ dominate and
these states give rise to diffusion processes against the
potential barriers $V_n^+(z)$. In the ultracold regime, this
explains why the atoms are reflected with high probability.

The tuning of the mazer from a perfect transmitter to a perfect
reflector system is a general property of the mazer when the
atom-field internal state is initially prepared in the trapping
state $|\gamma^+\rangle$. This property holds for any cavity field
mode and is independent of the cavity length (as it is based on
Eqs.~(\ref{RandTsimplified}) valid in these general conditions).
More restrictively, this property may also be observed using the
$|\gamma^-\rangle$ states as these states verify also the relation
(\ref{ratio}). In this situation, when $|\gamma|$ varies from 0 to
1, the wavefunction components over the states $|-,n\rangle$
increase and the particle is diffused more and more over the
potential wells $V_n^-(z)$. In the ultracold regime, this may
contribute to reflect as well the atoms, depending on the cavity
mode function. This is in particular the case for the mesa mode
function ($u(z) = 1$ inside the cavity and zero elsewhere) as it
is illustrated on Fig.~\ref{Figure1}. This figure represents the
reflection probability (\ref{R1}) with respect to $|\gamma|$ for
the $|\gamma^+\rangle$ and $|\gamma^-\rangle$ states. The
reflection coefficient $r_n^{\pm}(k)$ have been calculated using
the results of L\"offler \etal~\cite{Lof97}. The curve
corresponding to the $|\gamma^-\rangle$ state (denoted $R^-$ on
the figure) is specific to the mesa mode function. This is not the
case for the curve corresponding to the $|\gamma^+\rangle$ state
(denoted $R^+$) which, as expected, fits perfectly the general
result (\ref{Rsimplified}) valid for any cavity field mode.

\begin{figure}
\begin{center}
\noindent\fbox{\includegraphics[width=8cm, trim= 30 420 40 120,
keepaspectratio]{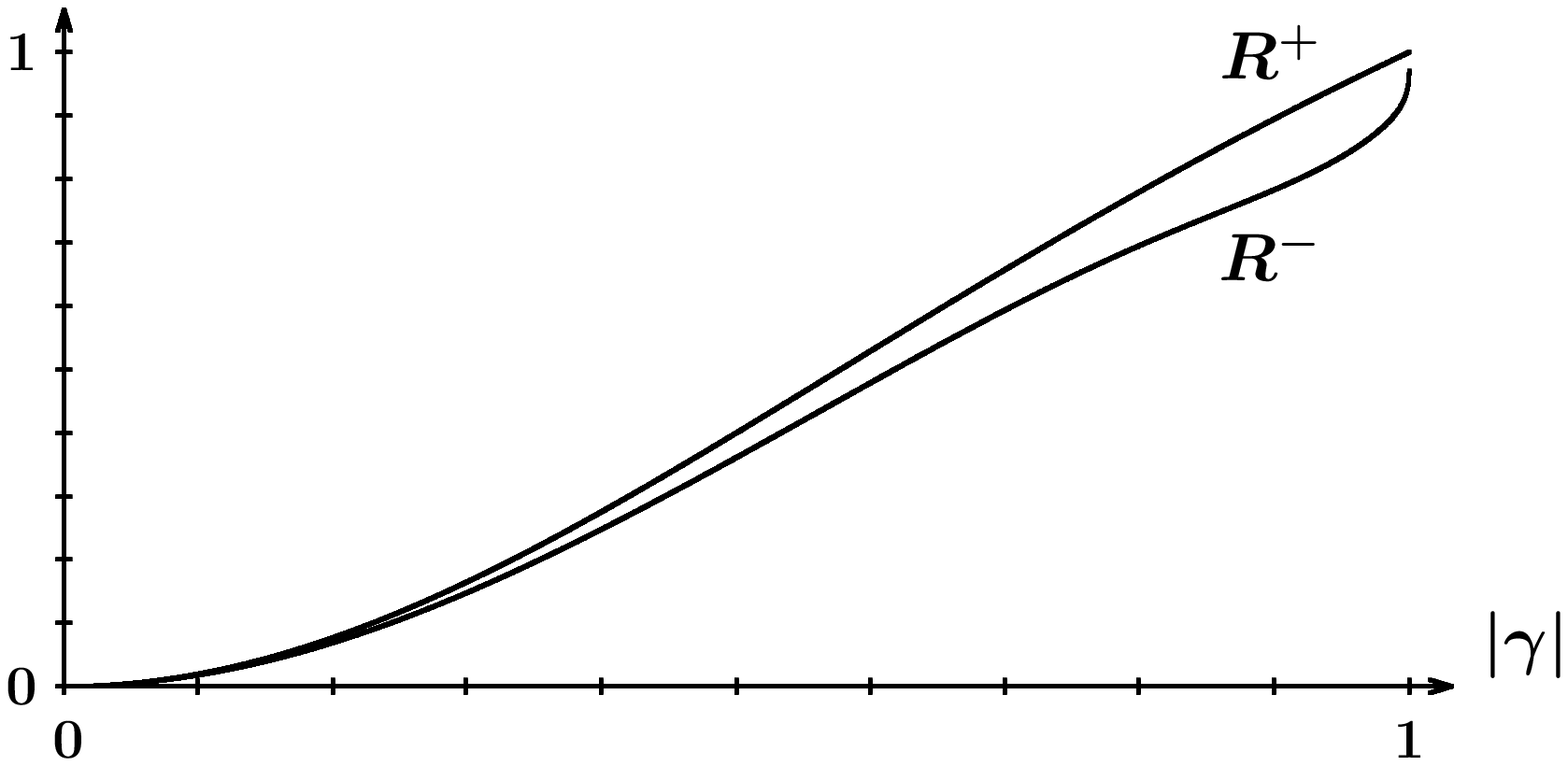}}
\end{center}
\caption{Reflection probabilities of the mazer in the
$|\gamma^+\rangle$ and $|\gamma^-\rangle$ states as a function of
$|\gamma|$ (respectively denoted $R^+$ and $R^-$). The cavity mode
profile is the mesa function, $k/\kappa = 0.1$ and $\kappa L =
10$.} \label{Figure1}
\end{figure}

\section{Experimental Issues}
\label{ExperimentalIssues}

The feasibility of a mazer-type experiment in the microwave or
optical domain has been discussed by L{\"o}ffler
\etal~\cite{Lof97} and Retamal \etal~\cite{Ret98}. High$-Q$
cavities are needed to fulfill the strong coupling condition and
avoid the spontaneous emission during the interaction of the atom
with the cavity. Also, kinetic energy significantly lower than the
interaction energy $\hbar g$ ($k/\kappa_n \ll 1$) are required to
be in the ultracold regime. Such experimental condition have been
achieved for the first time recently by Hood \etal~\cite{Hoo98} in
the optical domain (see also \cite{Hoo00,Pin00}). Nevertheless,
these results will have still to be improved to test the effects
presented in this work.

The population trapping condition, that preserves the internal
atomic state while keeping special scattering properties, are
attractive from an experimental point of view, since they do not
have any restriction about the cavity field mode, the cavity
interaction length and the initial atomic velocity. The fact that
these perfect trapping states present only one kind of potential,
a barrier or a well, to the incident ultracold atoms may be used
for new cavity QED scattering experiments \cite{Kae99}.

\section{Summary}
\label{SummarySection}

In this paper, we have studied the population trapping phenomenon
for the one-photon mazer. With the aim of such study in view, we
have written the quantum theory of the mazer by use of the
dressed-state coordinate formalism. Simple expressions for the
atomic populations, the cavity photon statistics, and the
reflection and transmission probabilities have been given for any
initial pure state of the atom-field system.

We have demonstrated that the population trapping phenomenon is
not only preserved in the ultracold regime but also exhibits new
properties. When the atom-field system is prepared in a perfect
trapping state, the scattering process becomes very particular.
Instead of ``feeling'' both a potential barrier and a potential
well, atoms passing through the cavity are only sensitive to one
of the potential components. Also, the atomic and field
populations are not changed while having the possibility of a
non-zero reflection coefficient.

\section{Acknowledgments}
\label{Acknowledgments} This work has been supported by the
Belgian Institut Interuniversitaire des Sciences Nucl\'eaires
(IISN) and partially by the Brazilian Conselho Nacional de
Desenvolvimento Científico (CNPq). E. S. wants to thank T. Bastin
for the hospitality at Universit{\'e} de Li{\`e}ge in Belgium.


\end{document}